# Nerve impulse propagation and wavelet theory

Louis Gaudart and Jean Gaudart*

Louis Gaudart, Jean Gaudart*
Aix Marseille Univ, IRD, INSERM, SESSTIM, Marseille, France

* Corresponding author.
SESSTIM UMR1252, Faculté de Médecine, 27 Bd Jean Moulin, 13385 Marseille cedex 05 France
Tel: + 33 491 791 910; Fax: + 33 491 794 013
*E-mail address*: jean.gaudart@univ-amu.fr

**Abstract**

A luminous stimulus which penetrates in a retina is converted to a nerve message. Ganglion cells give a response that may be approximated by a wavelet.

We determine a function $\Psi$ which is associated with the propagation of nerve impulses along an axon. Each kind of channel (inward and outward) may be open or closed, depending on the transmembrane potential. The transition between these states is a random event. Using quantum relations, we estimate the number of channels susceptible to switch between the closed and open states. Our quantum approach was first to calculate the energy level distribution in a channel. We obtain, for each kind of channel, the empty level density and the filled level density of the open and closed conformations. The joint density of levels provides the transition number between the closed and open conformations. The algebraic sum of inward and outward open channels is a function $\Psi$ of the normalized energy E. The function $\Psi$ verifies the major properties of a wavelet. We calculate the functional dependence of the axon membrane conductance with the transmembrane energy.

1. Introduction

The human vision possesses numerous specificities and the visual analysis of an image is achieved through a substantial efficiency. At first sight, the visual system shows a complex structure. A light stimulus which arrives on the eye passes through the eyeball. At the back of the eyeball the light beam arrives on the retina. Inside the retina the luminous stimulus is completely transformed. First the phototransduction of the stimulus into a nerve impulse is occurred and then the nerve message is coded by several layers of neuronal cells. Previous studies give one theory per visual mechanism. We investigated a unique theoretical model which is susceptible to take into account all the visual mechanisms, as color, contrasts, binocular vision. To model the visual mechanisms, we have considered the energy as basis variable.

The retinal circuitry is organized in a precise manner (Lee 1996; Lee and Dacey 1997; Kolb 1991). The retina is composed of three layers of nerve cells and synapses. The first layer contains the photoreceptor cells: rods and cones. These cells transduce the luminous energy into an electrical energy. The second layer is the outer plexiform layer and contains the horizontal cells. Some of coding begins within this layer. The third layer is the inner plexiform layer and contains the bipolar cells (diffuse-, midget-, S cone-bipolar cells…), the amacrine cells, and the ganglion cell terminals (parasol-, midget-, bistratified-cell terminations…). A more elaborate processing is occurred within this layer. The nerve message contains all the major characteristics of the light stimulus. The set of the ganglion axons constitutes the optical nerve which transmits the nervous message to the lateral geniculate nucleus, where a new coding is occurred and the binocular vision processing takes place. Similar occurrences are found at the cortical level with, for example, new receptive fields (Ts'o 1989; Ts'o and Gilbert 1988).

The color processing is a very complex processing (De Valois 1978; De Valois and De Valois 1993; De Valois and De Valois 1996). For a photopic vision and a normal subject, three wavelength intervals are isolated by three cone classes. The S-cone sensitivity is located in the short wavelengths, the M-cone sensitivity is located in the middle wavelengths, and the L-cone sensitivity is located in the long wavelengths. Two possibilities can be considered. The stimulus is either a small spot moving in direction or a larger stimulus having a fixed position, while its luminance is a function of time. The shape of the response is shown in figure 1.

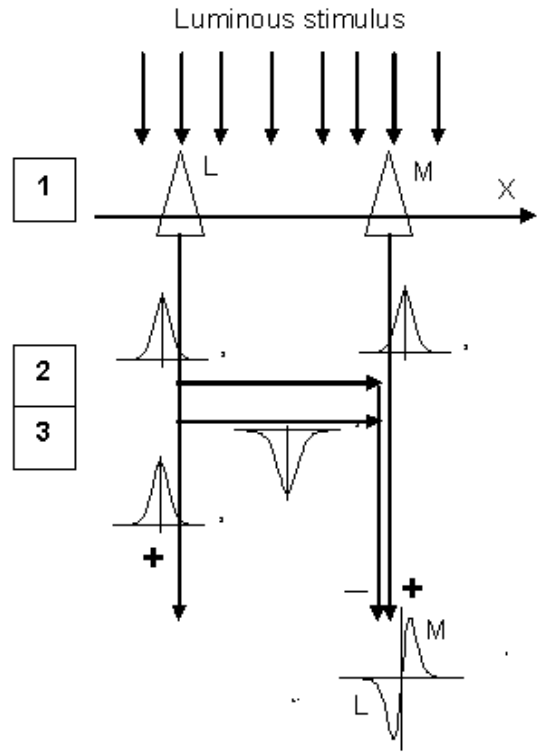

Fig. 1. Diagram model used to understand how a wavelet can be constructed. The direct pathways are constituted by the photoreceptors (M- and L-cone), the bipolar cells, and the ganglion cells. These pathways transmit the information without change (plus sign). The lateral pathways are constituted by the horizontal cells (2) and the amacrines cells (3). These pathways transmit the information with change in the sign (minus sign).

The influence of the horizontal and amacrines cells on the synapses of the bipolar and ganglion cells is complicated and not fully known (Wu et al. 2000; Pang et al. 2002). To establish a theoretical model, we have simplified the visual system, taking into account the main direct and lateral pathways. Two pathways are used to reach the optical nerve. There is a direct pathway including photoreceptor cells, bipolar cells, and ganglion cells. This pathway transmits the visual information without change in the shape of the theoretical curve. The lateral bindings between two or more cones of different types provide lateral transmissions and coding. Thus, a L-cone response can be associated with a M-cone response using horizontal cells (outer plexiform layer: Fig. 1, layer 2) or amacrines cells (inner plexiform layer: Fig. 1, layer 3). The lateral pathways transmit the visual information with a change in the shape of the theoretical curve, the sign of the response changes. At the ganglion axon, a direct pathway gives an "ON"-response, whereas an association of direct and lateral pathways gives an "OFF"-response. Thus, a receptive field having opponent centre-surround and spectrally-opponent information is obtained.

It is particularly interesting to note that some of the curves obtained inside the retina exhibit a wavelet shape. The role of a wavelet analysis in the human vision is an open question and a multiresolution decomposition seems a particularly interesting view. We attempt to investigate a simple and unique mother wavelet, which could be used as a basis to study all the visual mechanisms and eventually to study other physiological processes.

In a physiological system the information is transmitted along nerves with an "ON-OFF" signal. The simplest function to model an "ON-OFF" signal is the Haar function, defined by the following relations (Meyer 1994):

$$\begin{cases} H(x) = 0 & if \quad x < 0 \\ H(x) = 1 & if \quad 0 \leq x \leq 0.5 \\ H(x) = -1 & if \quad 0.5 \leq x \leq 1.0 \\ H(x) = 0 & if \quad x > 1.0 \end{cases}$$

The Haar function possesses the most properties of a wavelet. A wavelet family is built from a mother wavelet by including a dilation parameter "a" and a translation parameter. These parameters are of high interest to model the human vision. Indeed, the vision can view either large images or details. Movements of the eyeball produce movements of the vision axis and a translation of the retinal image. A wavelet analysis can reflect these properties.

From a physiological standpoint, the Haar wavelet is not an adequate model to reflect the visual system response. Indeed, this wavelet is not continuous, whereas the responses of the physiological systems are continuous, as the transmission of a signal by a neuron or by a receptive field (Fig. 1). A more appropriate wavelet should be determined.

We establish a function which can be a mother wavelet, from nervous impulse transmission and voltage-dependent channel properties. This function verifies the main properties of a wavelet, leading to a wavelet family definition. We propose to use this wavelet family to model the whole visual system, from retina to cortical level.

## 2. Theory

In this section we try to answer the question: Is it possible to establish a wavelet function from the physiological properties of the nervous message?

**Transmission of the nervous message**

To build a function $\Psi$ associated with the transmission of the nervous message along an axon, let us first outline the main mechanisms of the transmission. The nervous message is an action potential propagating along an axon. There are local changes of the membrane that let Na+ or K+ ions pass through (Pichon et al. 2004). Others ions, as Ca2+ or Cl−, can move too, but their action can be considered as negligible. The membrane contains voltage-dependent channels which may be open or closed, depending on local voltage value. The resulting ion current which passes through the membrane depends on the number of open inward and outward channels. To provide a generalized description we have replaced the transmembrane potential (measured in volt) by the equivalent energy ε (measured in electron-volt).

During the transmission there are ion exchanges between the inside and outside of the axon. An action potential is a temporary change of the plasma membrane permeability of the axon related to the amount of Na+ and K+ ions. An action potential transmission is an energy transmission. At a given point of an axon this energy changes the local permeability of the membrane. The change in the permeability is due to the change in the open channel number. Inward and outward channel openings play a major role in the nerve impulse transmission (Varshney and Mathew 2003). The main role of the energy variation is to open -or close- the voltage-dependent channels. The resulting ion current will be toward inside if the number of the open inward channels is larger than the number of the open outward channels. It will be toward outside if otherwise. This number is a function of the transmembrane potential and consequently of the corresponding energy $\varepsilon$. To switch between the closed and open states, the voltage-dependent channels should receive an energy. We study the probability of a channel opening at a given energy $\varepsilon$. The opening of a channel is a random event (Endresen et al. 2000). This means that we can use a quantum approach to calculate the probability of this event (Beck 1996).

We are particularly interested by two types of voltage-dependent channels: the inward channels $C_i$ (Na channels, for example) and the outward channels $C_O$ (K channels, for example). The conduction of ions through a channel is susceptible to be affected by the occurrence of others ions (Ca2+, for example; Horigan et al. 1999a; Horigan et al. 1999b). Moreover, Na channels allow a few numbers of K ions to pass (Immke et al. 1998, Kiss et al. 1998). To take into account the complex interactions between the various ions and channels, $C_i$ and $C_O$ are not exactly the same as the actual channels. We define $C_i$ and $C_O$ channels as equivalent to both actual channels and interactions. Each channel is regarded as having an individual equivalent charge. The equivalent channels $C_i$ and $C_O$ contain all the mechanisms that occur in an actual channel. Consequently, the following relations can be used, if need be, in the case of other type of voltage-dependent channels.

A channel is made of charged molecules (Bashford 2004; Chung and Kuyucak 2002). The voltage across the membrane has an influence on these charges. When a channel switches between the closed and open states, the charges move. The energy configuration of the channel is modified. Each type of channels has two main energy configurations: a closed and open configuration. The energy configuration depends on the transmembrane potential, namely the transmembrane energy. Recent physiological studies exhibit several intermediate states. Intermediate states are induced by movements of specific charges located on the channel molecules (Elinder et al. 2001). To establish a quantum model we consider an initial state, the closed state, and a final state, the open state. The energy variations between these states take into account all the intermediate states. A channel opening is induced by specific charge movements (Horn 1997). These charges are called gating charges. The charge movements constitute the gating current (Sigg and Bezanilla 1997). The movements of the gating charges are random events and it is beneficial to use a quantum approach. The charge movements are induced by energy transfers that modify the state of the voltage-dependent channel. We define one equivalent theoretical charge per channel. An equivalent theoretical charge is constituted by the set of the actual charges required during a channel opening. Thus, an equivalent charge could be constituted by various charges: effective charges, essential charges, peripheral charges and latent charges (Bezanilla 2000). The equivalent charge is a convenient theoretical model to calculate the closed-open transition probability. We study the probability of the equivalent charge configuration changes *i.e.* the probability of a voltage-dependent channel opening.

Each configuration, closed or open, possesses filled energy levels (bottom of the energy diagram) and above empty energy levels (Fig. 2). The energy level distribution is continuous. These two configurations are separated by an energy barrier and are stable (Liebovitch and Todorov 1996). An equivalent charge has a probability $\Phi(\varepsilon)$ to locate on a filled level having an energy $\varepsilon$. In the open state the distribution is similar, with filled energy levels and empty energy levels. If there are no empty levels that can be filled, then, there is no transition possible between the closed state and the open one. Moreover, the channel transition from the closed state to the open state can occur only if the upper filled level $\varepsilon_1$ of the closed state is above the upper filled level $\varepsilon_2$ of the open state.

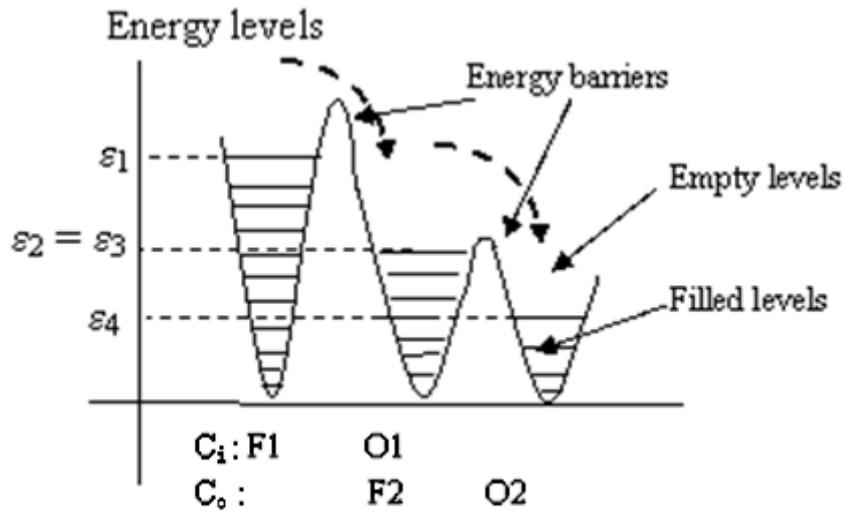

Fig. 2. Closed-open transitions. An energy $\Delta\varepsilon$ arriving in a closed channel causes energy barrier decay. F1: energy levels of an equivalent charge of a closed inward channel $C_i$. O1: energy levels of an equivalent charge of an open inward channel $C_i$. F2: energy levels of an equivalent charge of a closed outward channel $C_o$. F2 is the same that O1. O2: energy levels of an equivalent charge of an open outward channel $C_o$.

Schematically we have:

A local energy Δε arrives
⇓
The energy barrier decreases
⇓
The upper filled levels of the closed state empty
⇓
Some of empty levels of the open state fill
⇓
The channel opens and Δε is emitted
⇓
The energy Δε is transmitted to a neighboring closed channel

The switching between the closed and open state is a random event (Liebovitch and Todorov 1996). The switching probability between the states can be calculated. Thus the mean number of closed-open transitions can be determined at a given energy ε. This number depends on both the distribution of the filled levels that can empty and distribution of the empty levels that can fill.

The channels are made up of similar proteins. The distribution of the empty and filled levels is the same in the inward and outward channels but the upper filled level is not the same. To establish a theoretical model, we consider two major features of the channel openings. First, $C_i$ channels open and Na ions pass inside the membrane. The transmembrane energy increases and reaches a maximal value. Second, the $C_o$ channels open and K ions flow out the membrane. These steps are described in Fig. 2. The arrival of energy Δε causes a barrier potential decrease. The filled levels of the closed inward channel empty and the empty levels of the open state fill, until an intermediate value $ε_2$. The inward channel opens. Filled levels of the closed outward channel empty and empty levels of the open outward channel fill.

The various steps of an action potential are the same that the energy variation of our model.

$\varepsilon_1 \to \varepsilon_2$ : inward channel $C_i$ opening

$\varepsilon_2 = \varepsilon_3 \to \varepsilon_4$ : outward channel $C_o$ opening

$\Phi(\varepsilon-\varepsilon_1)$, $\Phi(\varepsilon-\varepsilon_2)$, $\Phi(\varepsilon-\varepsilon_4)$ probability of an equivalent charge to have an energy $\varepsilon$

$\Phi_e(\varepsilon-\varepsilon_1)$, $\Phi_e(\varepsilon-\varepsilon_2)$, $\Phi_e(\varepsilon-\varepsilon_4)$ probability of an empty level.

The relative position of the upper levels is important, and the mechanism of the nervous message transmission is based on this position.

**Probability of closed-open transitions**

The probability of a channel opening is given by both the filled level distribution and empty level distribution. The probability $\Phi(c \to o)$ of a closed-open transition is given by the joint density of the quantum levels. In each state, closed or open, the equivalent charge has a probability $\Phi(\varepsilon)$ to be on an energy level $\varepsilon$.

The probability of a closed inward channel (upper level $\varepsilon_1$) having a filled level at energy $\varepsilon$ (Fig. 3) is:

$$\Phi_{ci}(\varepsilon - \varepsilon_1) = \frac{1}{1 + e^{\frac{\varepsilon - \varepsilon_1}{BT}}} = \left(1 + e^{\frac{\varepsilon - \varepsilon_1}{BT}}\right)^{-1} \qquad (1)$$

The probability of an open inward channel (upper level $\varepsilon_2$) having a filled level at energy $\varepsilon$ is:

$$\Phi_{oi}(\varepsilon - \varepsilon_2) = \frac{1}{1 + e^{\frac{\varepsilon - \varepsilon_2}{BT}}} = \left(1 + e^{\frac{\varepsilon - \varepsilon_2}{BT}}\right)^{-1} \qquad (2)$$

B is the Boltzmann constant; T is the absolute temperature.

If T ≠ 0 K the distribution is continuous with a probability $\Phi(\varepsilon)$ to find an equivalent charge above the energy $\varepsilon_1$ or $\varepsilon_2$.

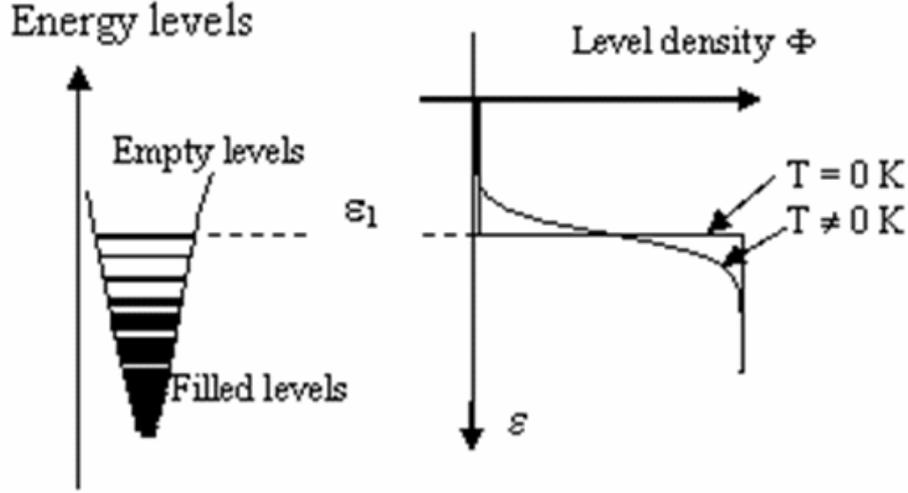

Fig. 3. Energy level distribution. At the temperature T = 0 K, the energy $\varepsilon_1$ is the border between the filled levels and the empty ones. Then the level density curve exhibits a step at $\varepsilon_1$. If the temperature is T ≠ 0 K, there are some filled levels above this energy.

A voltage-dependent channel can switch between a closed and open state if the following condition is verified: $\varepsilon_1 > \varepsilon_2$

In the case of an open state we need an empty level distribution, not a filled level distribution. The empty level distribution $\Phi_{oie}$ of an inward channel having an open configuration is given by:

$$\Phi_{oie}(\varepsilon - \varepsilon_2) = 1 - \Phi_{oi}(\varepsilon - \varepsilon_2) = 1 - \left(1 + e^{\frac{\varepsilon - \varepsilon_2}{BT}}\right)^{-1} \qquad (3)$$

A transition needs to have a filled level ($\Phi(\varepsilon-\varepsilon_1) \neq 0$) associated with an empty level ($\Phi_e(\varepsilon-\varepsilon_2) \neq 0$). The probability to switch between a closed state (Fig. 4, F1 or F2) and an open state (O1 or O2) depends on both filled levels of F1 (or F2) and empty levels of O1 (or O2). With high values of energy $\varepsilon$ we have $\Phi(\varepsilon-\varepsilon_1) = 0$ (Fig. 4, state F1, filled levels, upper curve) and there is no filled level. Consequently, no transition is possible. With small values of energy ε, $\Phi_e(\varepsilon-\varepsilon_2) = 0$ (Fig. 4, state O1, empty levels, dashed curve), there is no empty level. In this case the energy ε is located in the filled band of the open state. No transition can occur toward these levels that are already filled.

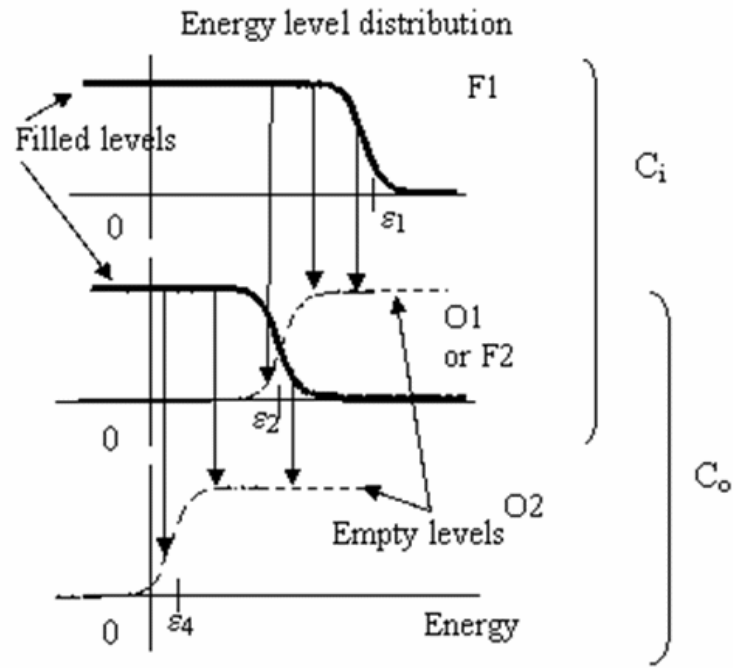

Fig. 4. Continuous distribution of the energy levels at $T \neq 0$ K. $C_i$: inward channel; $C_o$: outward channel. The abscissa values represent the energy ε. The ordinate values represent the energy level distribution. $\varepsilon_1$, $\varepsilon_2$ and $\varepsilon_4$ is the energy of the upper filled level (at $T = 0$ K) of the closed inward channel, open inward channel (or closed outward channel) and open outward channel, respectively. F1: closed inward channel. O1: open inward channel. F2: closed outward channel. The energy distribution of O1 and F2 is the same. O2: open outward channel. The vertical arrows indicate closed-open transitions between two levels.

An equivalent charge located of one filled level of the closed state (probability $\Phi_{ci}(\varepsilon-\varepsilon_1)$) can fall down all the empty levels of the open state (probability: $\Phi_e(\varepsilon-\varepsilon_2)$). From one filled energy $\varepsilon$ the transition probability is: $\Phi_e(\varepsilon-\varepsilon_2)$. If we take into account the total filled levels of the closed state and the total empty levels of the open states, then all the possible transitions are obtained. Closed-open transition of an inward channel $C_i$:

F1($\varepsilon_1$) $\rightarrow$ O1($\varepsilon_2$); energy variation: $\varepsilon_2 - \varepsilon_1$

Closed-open transition of an outward channel $C_o$:

F2($\varepsilon_3 = \varepsilon_2$) $\rightarrow$ O2($\varepsilon_4$); energy variation: $\varepsilon_4 - \varepsilon_2$

The transition probability is the product of the filled level distribution of the closed state times the empty level distribution of the open state. The product is the joint density of the quantum states. The joint density of the quantum states of an inward channel is given by:

$$\Phi_i(c \rightarrow o) = \Phi(\varepsilon-\varepsilon_1) \times \Phi_e(\varepsilon-\varepsilon_2) = \Phi(\varepsilon-\varepsilon_1) \times \left[1-\Phi(\varepsilon-\varepsilon_2)\right] \tag{4}$$

The joint density of the quantum states of an outward channel is given by:

$$\Phi_o(c \rightarrow o) = \Phi(\varepsilon-\varepsilon_2) \times \Phi_e(\varepsilon-\varepsilon_4) = \Phi(\varepsilon-\varepsilon_2) \times \left[1-\Phi(\varepsilon-\varepsilon_4)\right] \tag{5}$$

These equations give the closed-open transition probability of a voltage-dependent channel, namely the probability of a voltage-dependent channel to be open at an energy $\varepsilon$. If the voltage-dependent channel number per unit surface of membrane is known, then these equations give the number of open channel. In the following, we will consider $\Psi$ as a normalized number of open voltage-dependent channel.

An action of an inward channel on the ion current is opposed by an action of an outward channel. The resulting amount of open channels that let ions pass inside the membrane is:

$\Psi(\varepsilon) = \Psi_i(\varepsilon) - \Psi_o(\varepsilon)$

$\Psi_i(\varepsilon)$ is the normalized number of open inward channels.

$\Psi_o(\varepsilon)$ is the normalized number of open outward channels.

If $\Psi(\varepsilon) > 0$ then the number of open inward channels is greater than the number of open outward channels. The ion flux passes inside the membrane.

If $\Psi(\varepsilon) < 0$ then the number of open inward channels is less than the number of open outward channels. The ion flux moves outward through the membrane.

### 3. Results

**Calculations of the function $\Psi$**

The basis of the relationship between the normalized number $\Psi$ of open channels and the energy $\varepsilon$ is: $\Psi(\varepsilon) = \Psi_i(\varepsilon) - \Psi_o(\varepsilon)$

$$\Psi(\varepsilon) = \left(1 + e^{\frac{\varepsilon - \varepsilon_1}{BT}}\right)^{-1} \times \left[1 - \left(1 + e^{\frac{\varepsilon - \varepsilon_2}{BT}}\right)^{-1}\right] - \left(1 + e^{\frac{\varepsilon - \varepsilon_2}{BT}}\right)^{-1} \times \left[1 - \left(1 + e^{\frac{\varepsilon - \varepsilon_4}{BT}}\right)^{-1}\right] \quad (6)$$

To simplify, let us introduce:

$$u_1 = \frac{\varepsilon_1 - \varepsilon_2}{BT} \qquad u_2 = \frac{\varepsilon_4 - \varepsilon_2}{BT} \qquad u = \frac{\varepsilon - \varepsilon_2}{BT}$$

The channels are made up of similar proteins. The difference between the upper level of the closed state and the upper level of the open state is similar in both inward and outward channel. Consequently, we have:

$$(\varepsilon_1 - \varepsilon_2) = (\varepsilon_2 - \varepsilon_4) \Leftrightarrow \varepsilon_2 = \frac{\varepsilon_1 + \varepsilon_4}{2} \tag{7}$$

$$u_1 + u_2 = \frac{1}{BT}[(\varepsilon_1 + \varepsilon_4) - 2\varepsilon_2] = 0$$

$$u_1 = -u_2$$

$$\frac{\varepsilon - \varepsilon_1}{BT} = u - u_1 = u + u_2; \quad \frac{\varepsilon - \varepsilon_4}{BT} = u - u_2$$

Then we obtain:

$$(\text{Eq. 6}) \Rightarrow \Psi(u) = \frac{e^u[(1-e^u)(1-e^{u_2})]}{(1+e^{u+u_2})(1+e^u)(1+e^{u-u_2})} \tag{8}$$

To simplify the relation and exhibit symmetry we use hyperbolic functions.

$$\tanh(-\frac{u}{2}) = \frac{1-e^u}{1+e^u}; \quad \coth(-\frac{u}{2}) = \frac{1}{\tanh(-\frac{u}{2})}$$

Then:

$$\Psi(u) = \frac{e^{-u_2}}{1+e^{-u_2}} \times \frac{1 - \tanh^2(-\frac{u}{2})}{\coth(\frac{u_2}{2})\coth(-\frac{u}{2}) - \tanh(\frac{u_2}{2})\tanh(-\frac{u}{2})}$$

We define:

$$A = \frac{e^{-u_2}}{1+e^{-u_2}} \qquad E = -\frac{u}{2} = -\frac{\varepsilon - \varepsilon_2}{2BT} \qquad E_2 = \frac{u_2}{2} = -\frac{u_1}{2} = -\frac{\varepsilon_1 - \varepsilon_2}{2BT}$$

The constants A and $E_2$ depend on the energy difference that allows a voltage-dependent channel opens. The function $\Psi$ depends on energy differences. Consequently, it is not necessary to know the absolute value of the energy. The variable E is a normalized energy.

The following relation is obtained:

$$\Psi(E) = A \times \frac{1-\tanh^2(E)}{\coth(E_2)\coth(E) - \tanh(E_2)\tanh(E)} \qquad (9)$$

The function Ψ(E) is associated with the nervous message transmission. It represents the normalized difference between the open inward channel number and open outward channel number, namely the resulting number of open channels. It also indicates the direction of the global ion flux.

We have determined A and $E_2$ from the existing data.

$B = 0.862 \times 10^{-4}$ eV×K$^{-1}$ ; T = 300 K

$\varepsilon_1 - \varepsilon_2 = 0.050$ eV

$$E_2 = -\frac{\varepsilon_1 - \varepsilon_2}{2BT} = -0.97 \cong -1.0$$

$$\Psi(E) = 0.87 \times \frac{1 - \tanh^2(E)}{-1.34\coth(E) + 0.75\tanh(E)}$$

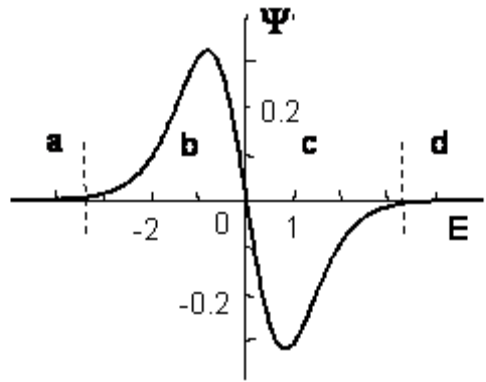

Fig. 5. Mean number of open equivalent channels Ψ as a function of normalized transmembrane energy E. As Ψ is positive (part "b"), then the open inward channel number is larger than the open outward channel number. The total ion flux is inwardly. As Ψ is negative (part "c"), the total ion flux moves outward through the membrane.

Several main energy intervals can occur (Fig. 5).

$\varepsilon$ is very large:  $\varepsilon_4 < \varepsilon_2 < \varepsilon_1 < \varepsilon$ ;   $E < 0$

$\varepsilon$ is located in empty levels of F1 (Fig. 4). There is no transition (Fig. 5, part « a »).

$\varepsilon_4 < \varepsilon_2 < \varepsilon < \varepsilon_1$                    $E < 0$

$\varepsilon$ is located in filled levels of F1 and the corresponding levels of O1 are empty (Fig. 4). There is a transition. The number of open channels depends on the position of $\varepsilon$ between $\varepsilon_1$ and $\varepsilon_2$. The inward channels can open (Fig. 5, part "b"). The decaying phase of $\Psi$ indicates that the outward channel openings start.

$\varepsilon_4 < \varepsilon < \varepsilon_2 < \varepsilon_1$                    $E > 0$

$\varepsilon$ is located in filled levels of F1 and the corresponding levels of O1 is filled too (Fig. 4). The inward channel cannot open. However, since $\varepsilon$ is larger than $\varepsilon_4$ the corresponding levels of F2 are filled and the corresponding levels of O2 are empty. The outward channels can open (Fig. 5, part "c"). The number of open outward channels is larger than the number of open inward channels.

$\varepsilon$ is very small:  $\varepsilon < \varepsilon_4 < \varepsilon_2 < \varepsilon_1$          $E > 0$

$\varepsilon$ is located in filled levels of both F1 and O1, F2 and O2 (Fig. 4). No transition can occur (Fig. 5, part "d").

The shape of the curve is similar to the response of the retina cells (Fig. 1).

Usually, the probability or the mean number of open voltage-dependent channels is given as a function of time (Horn et al. 1981). In this study we have made an energy balance. It is important to use energy as fundamental variable to provide a full description of the transmission of a stimulus that is susceptible to be transduced and coded.

**Properties of the function Ψ: is the function Ψ a wavelet function?**

In this study we consider the action potential as a signal which is transmitted inside the nervous system. Consequently, we hope to benefit from this to consider the wavelet theory. A function is a basic wavelet if it verifies several main conditions (Daubechies 1992; Meyer 1994; Holschneider 1995).

The function Ψ verifies the following conditions.

    *a) The function integral converges:*

$$\int_{-\infty}^{+\infty} \Psi(E) dE < \infty$$

    *b) The function integral is null and the condition for zero mean is:*

$$\int_{-\infty}^{+\infty} \Psi(E) dE = 0$$

The curve exhibits oscillations with negative part and positive one. The curve is a small wave.

    *c) The Fourier transform of a wavelet is null when $\omega = 0$.*

If $\omega = 0$ then: $\int_{-\infty}^{+\infty} \Psi(E) e^{-i\omega E} dE = 0$

    *d) The function has a higher number of vanishing moments.*

1 order moment $\quad \int_{-\infty}^{+\infty} E \times \Psi(E) dE = 0$

2 order moment $\quad \int_{-\infty}^{+\infty} E^2 \times \Psi(E) dE = 0$

All the even moments are null.

    e) $\lim_{|E_2| \to \infty} (\Psi) = Haar\ function$

The normalization condition is given by:

$$\int_{-\infty}^{+\infty} |\Psi_N(E)|^2 dE = 1 \qquad \Psi_N \text{ is a normalized wavelet.}$$

The normalized mother wavelet is:

$$\Psi_N = \frac{1}{\sqrt{N}} \Psi(E) \qquad \text{the normalization factor is: } N = \int_{-\infty}^{+\infty} |\Psi(E)|^2 dE$$

Consequently, the function $\Psi$ is a wavelet.

The wavelet function contains a parameter $E_2$ which denotes a physiological condition (Fig. 6). It depends on characteristics of voltage-dependent channels.

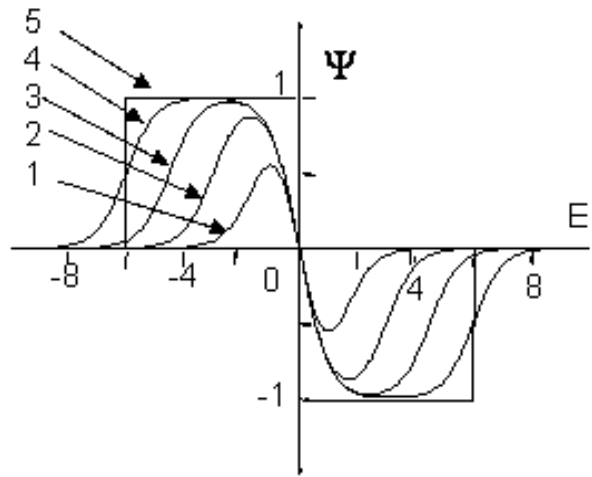

Fig. 6. Variation in wavelet with parameter value $E_2$. Curve 1: $E_2 = -1.5$; curve 2: $E_2 = -3$; curve 3: $E_2 = -4.5$; curve 4: $E_2 = -6$; curve 5: Haar function.

## 4. Discussion

The transmembrane conductance is given by the number of open inward channels. One test of the theoretical model is to deduce the conductance curves. If the theoretical model is adequate, then the theoretical conductance curves should be similar to the experimental curves. The validity of the model will be confirmed.

At an energy ε the electrical conductance of the membrane is given by the number of all open channels in an interval less than ε eV. The local conduction C(ε) of an axon membrane is given by:

$$C(\varepsilon) = \int_{-\infty}^{\varepsilon} \Psi_i(\varepsilon) d\varepsilon \qquad (10)$$

Corresponding curves are plotted Fig. 7. The curves are similar to those obtained by several authors (Bezanilla 2000; Clay 2000; Sigg and Bezanilla 1997; Varshney and Mathew 2003).

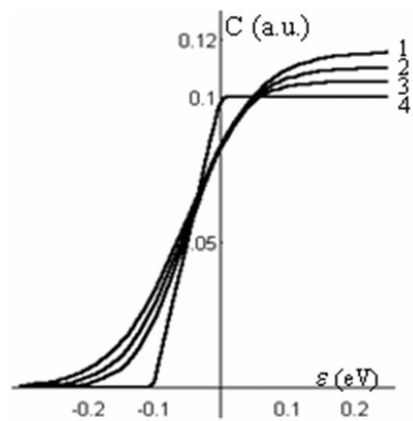

Fig. 7. Transmembrane energy dependence of conductance. T = 300 K.

Curve 1: $E_2 = -1$; curve 2: $E_2 = -1.2$; curve 3: $E_2 = -1.5$; curve 4: $E_2 = -16$.

Let us outline that three major physiological results correspond to three wavelet properties. First, an opening of channel is a random event. The constant field model (Goldman-Hodkin-Katz approximations) is not appropriate at the microscopic level (Syganow and von Kitzing 1999). We use a quantum approach to determine the probability of this event.

Second, the wavelet function is the result of two channel openings, inward channel and outward channel open successively. The energy Δε involved in the closed-open transition is always the same. It is equal to a difference between two energy levels and consequently independent of the incident stimulus. The energy which propagates along an axon is equal to the energy difference Δε which is constant. This theoretical result is similar to the physiological data at which the height of an action potential is induced by an energy difference and it is a constant for a given nervous cell.

Third, schematically, information is transmitted along an axon using an ON-OFF mechanism. A wavelet (function integral is null) is an adequate function to model this basis property.

A wavelet function is especially attractive in the signal theory mainly because a wavelet transform can be constructed (Mallat 1989a; Mallat 1989b). This transform gives the response of a system that receives an incident signal.

If the nervous fibre receives an outer stimulus, then the basic nervous message is modified. Let us consider an outer stimulus S(E). The S(E) transform by the wavelet Ψ(E) is:

$$W.T[S(E)] = \int_{-\infty}^{+\infty} S(E) \times \Psi(E) dE$$

The stimulus modifies the voltage-dependent channel energy and then modifies the probability to have a channel opening. This change is given by W.T[S(E)].

The present wavelet analysis can be used to study the process of a luminous stimulus by the nervous system (Antoine et al. 1992). The shape of the wavelet is similar to the response of the retina cells (Fig. 1). This wavelet has been already used to model contrast sensitivity functions (Gaudart et al. 1993). Moreover, the wavelet constitutes a unique theoretical model which is able to models all visual mechanisms such as color, contrast, or binocular vision.

# 5. References

Antoine JP, Murenzi R, Piette B, Duval-Destin M (1992) Image analysis with 2D continuous transform: detection of position, orientation, and visual contrast of simple objects. In: Meyer Y (ed) Wavelets and applications. Masson, Paris, pp. 144-159

Bashford CL (2004) Ion permeation of pores in model membranes selectivity, fluctuations and the role of surface charge. Eur Biophys J 33: 280-282

Beck F (1996) Can quantum processes control synaptic emission? Int J Neural Systems 7: 343-353

Bezanilla F (2000) The voltage sensor in voltage-dependent ion channels. Physiol Rev 80: 555-592

Chung SH, Kuyucak S (2002) Ion channels: recent progress and prospects. Eur Biophys J 31: 283-293

Clay JR (2000) Determining K+ channel activation curves from K+ channel currents. Eur Biophys J 29: 555-557

Daubechies I (1992) Ten lectures on wavelets. Society Industrial and Applied Mathematics Philadelphia, Pennsylvania

De Valois KK (1978) Interactions among spatial frequency channels in the human visual system. In: Cool SJ and Smith EL (ed) Frontiers in Visual Science. Springer-Verlag New York, pp. 277-285

De Valois RL, De Valois KK (1993) A multistage color model. Vision Res 33: 1053-1065

De Valois RL, De Valois KK (1996) On « A three-stage color model ». Vision Res 36: 833-836

Elinder F, Arhem P, Larsson P (2001) Localization of the extracellular end of the voltage sensor S4 in a potassium channel. Biophys J 80: 1802-1809

Endresen LP, Hall K, Hoye JS, Myrheim J (200) A theory for the membrane potential of living cells. Eur Biophys J 29: 90-103

Gaudart L, Crébassa J, Pétrakian JP (1993) Wavelet transform in human visual channels. Applied Optics 32: 4119-4127

Holschneider M (1995) Wavelets. An analysis tool. Mathematical Monograph. Oxford University Press

Horn R (1997) Counting charges. J Gen Physiol 110: 129-132

Horigan FT, Cui J, Aldrich RW (1999a) Allosteric voltage gating of potassium channels I : mSlo ionic currents in the absence of Ca2+. J Gen Physiol 114: 277-304

Horigan FT, Aldrich RW (1999b) Allosteric voltage gating of potassium channels II : mSlo channel gating charge movement in the absence of Ca2+. J Gen Physiol 114: 305-336

Immke D, Kiss L, LoTurco J, Korn SJ (1998) Influence of non-P region domains on selectivity filter properties in voltage-gated K+ channels. Receptors and channels 6: 179-188

Kiss L, Immke D, LoTurco J, Korn SJ (1998) The interaction of $Na^+$ and $K^+$ in voltage-gated potassium channels. Evidence for cation binding sites of different affinity. J Gen Physiol 111: 195-206

Kolb H (1991) Anatomical pathways for color vision in the human retina. Visual Neuroscience 7: 61-74

Lee BB (1996) Receptive field structure in the primate retina. Vision Res 36: 631-644

Lee BB, Dacey DM (1997) Structure and function in primate retina. In: Cavonius CR (ed) Color Vision Deficiencies, vol 13. Kluwer Academic, London, pp. 107-117

Liebovitch LS, Todorov AT (1996) What causes ion channel proteins to fluctuate open and closed? Int J Neural Systems 7: 321-331

Mallat SG (1989a) A theory for multiresolution signal decomposition: the wavelet representation. I.E.E.E. Transaction on pattern analysis and machine intelligence 11: 674-693

Mallat SG (1989b) Multifrequency channel decompositions of images and wavelet models. I.E.E.E. Transaction on acoustic, speech and signal processing 37: 2091-2110

Meyer Y (1994) Les ondelettes. Armand Colin, Paris

Pang JJ, Gao F, Wu SM (2002) Relative contributions of bipolar cell and amacrine cell inputs to light responses of ON, OFF and ON-OFF retinal ganglion cells. Vision Res 42: 19-27

Pichon Y, Prime L, Benquet P, Tiaho F (2004) Some aspects of the physiological role of ion channels in the nervous system. Eur Biophys J 33: 211-226

Sigg D, Bezanilla F (1997) Total charge movement per channel: the relation between gating charge displacement and the voltage sensitivity of activation. J Gen Physiol 109: 27-39

Syganow A, von Kitzing E (1999) (In)validity of the constant field and constant currents assumptions in theories of ion transport. Biophys J 76: 768-781

Ts'o DY (1989) The functional organization and connectivity of color processing. In: Man-Kit L and Gilbert CD (ed) Neural mechanisms of visual perception. Gulf publishing company USA, pp 87-115

Ts'o DY, Gilbert CD (1988) The organisation of chromatic and spatial interactions in the primate striate cortex. J Neurosci 8: 1712-1727

Varshney A, Mathew MK (2003) Inward and outward potassium currents through the same chimeric human Kv channel. Eur Biophys J 32:113-121

Wu SM, Gao F and Maple BR (2000) Functional architecture of synapses in the inner retina: segregation of visual signals by stratification of bipolar cell axon terminals. J Neurosci 15: 4462-4470